# Discrimination of neutrons and γ-rays in liquid scintillator based on Elman neural network

**Cai-Xun Zhang(张才勋)[1]**    **Shin-Ted Lin(林兴德)[2]**    **Jian-Ling Zhao(赵建玲)[1]**
**Xun-Zhen Yu(余训臻)[2]**       **Li Wang(王力)[3]**          **Jing-Jun Zhu(朱敬军)[1:1]**
**Hao-Yang Xing(幸浩洋)[2:2]**

[1] Key Laboratory of Radiation Physics and Technology of Ministry of Education,
Institute of Nuclear Science and Technology, Sichuan University, Chengdu, 610065, China
[2] School of Physical Science and Technology, Sichuan University, Chengdu, 610065, China
[3] Key Laboratory of Particle and Radiation Imaging (Ministry of Education) and
Department of Engineering Physics, Tsinghua University, Beijing, 100084, China

∗ Supported by National Natural Science Foundation of China(11275134,11475117)

1) E-mail: zhujingjun@scu.edu.cn

2) E-mail:xhy@scu.edu.cn

# Abstract

**Abstract:** In this work, a new neutron and γ (n/γ) discrimination method based on an Elman Neural Network (ENN) is proposed to improve the discrimination performance of liquid scintillator (LS) detectors. Neutron and γ data were acquired from an EJ-335 LS detector, which was exposed in a $^{241}$Am-$^{9}$Be radiation field. Neutron and γ events were discriminated using two methods of artificial neural network including the ENN and a typical Back Propagation Neural Network (BPNN) as a control. The results show that the two methods have different n/γ discrimination performances. Compared to the BPNN, the ENN provides an improved of Figure of Merit (FOM) in n/γ discrimination. The FOM increases from 0.907 ±0.034 to 0.953 ±0.037 by using the new method of the ENN. The proposed n/γ discrimination method based on ENN provides a new choice of pulse shape discrimination in neutron detection.

**Keywords**: Liquid Scintillator; n/γ discrimination; Elman neural network; BP neural network
**PACS**: 29.25.Dz, 29.50.+v, 07.05.Mh

## 1. Introduction

The technique of neutron detection is important in basic physics research, especially in the direct detection of dark matter. Since the pulse shape of nuclear recoil signals induced by WIMPs (Weakly Interactive Massive Particles), which is supposed to be the most promising candidate for dark matter are the same as the pulse shape of signals induced by neutrons, understanding the neutron background in the environment is significant for dark matter research [1-4]. Furthermore, neutron background is always accompanied by γ radiation [5], thus, n/γ discrimination plays a key role in neutron detection. Because of its excellent discrimination capabilities and fast time response abilities, liquid scintillator (LS) is widely used for fast neutron detection. The key feature of LS for fast neutron detection in the presence of γ rays is that it shows slow components depending on the specific energy loss density of the ionizing particle. The pulse shape of signal induced by neutrons always has a high proportion of slow component while γ signals show the opposite effect. The difference of pulse shape of these two kinds of signals can be used to discriminate neutron and γ signals. Up to now, there have been many n/γ discrimination methods, such as the Charge Comparison Method (CCM) [6-7] and the Rise Time Method [8-9], which are based on analog electronic technology and complicated circuit configurations. Although these methods have excellent online data analysis properties, their stability is not as good as expected. Recently, with the development of high speed ADCs, digital signal processors (DSP) and field-programmable gate arrays (FPGA), it is possible to record pulse shape of signals generated in LS completely. These new technologies prompt the study of new methods of n/γ discrimination such as fuzzy c-means algorithm [10], wavelet transform [11], power spectrum gradient analysis [12] and support vector machine (SVM) [13]. As a major method of non-linear science and computational intelligence science, neural networks have been widely applied in pattern recognition, artificial intelligence and biological information. In 1998, Zhong et al. [14] first applied artificial neural networks to n/γ discrimination firstly. Then Esposito et al [15] and Liu et al [16] further developed the application of neural networks in this field. However, in these studies the neural network structures were all based on traditional back propagation neural networks (BPNN), which limits the use of dynamic information of data in neural networks. According to the previous research, the feedback Elman neural network (ENN) [17] has an excellent ability to process dynamic time-varying signals. Thus in this work an ENN is proposed to perform n/γ discrimination combining with the randomness and time-varying characteristics of nuclear signals.

## 2. ENN

The ENN is a dynamic regression neural network. It can map dynamic characteristics through storing the internal status, thus, it is able to adapt to signals with time-varying characteristics. Generally the ENN is composed of 4 layers: input layer, hidden layer, output layer and a specific context layer which is an additional layer for the traditional BPNN. The structure of an ENN is shown in Fig. 1. The input layer acts as signal transmission. The output layer plays the role of linear weighting. The hidden layer is a typical nonlinear activation function. The transfer function can be a linear or nonlinear function. The output value of training of the hidden layer is stored in the context layer through recursive connection. In the next training, the stored value in the context layer will work as feedback and be input into the hidden layer to affect the training process. The feedback process provides the network dynamic memory property. The feedback process is expressed as follows:

$$y(k) = g(w^3 x(k)), \tag{1}$$

$$x(k) = f(w^1 x_c(k) + w^2(u(k-1))),\qquad(2)$$

$$x_c(k) = x(k-1),\qquad(3)$$

where $y$ is the output layer node vector, $x$ is the hidden layer node vector, $u$ is the input layer vector, $x_c$ is the feedback vector, $W^1$ is the connection weight of the context layer and the hidden layer, $W^2$ is the connection weight of the input layer and the hidden layer, $W^3$ is the connection weights of the hidden layer and the output layer, $g()$ is the transfer function of the output neuron and $f()$ is the transfer function of the hidden layer neuron. The classification performance after this machine learning method can be evaluated using the following equation.

$$E(w) = \sum_{k=1}^{n}[yk(w) - y\bar{k}(w)]^2,\qquad(4)$$

where $yk(w)$ is the result calculated by the neural network and $y\bar{k}(w)$ is the training data set. The idea of this algorithm here is as following. In order to get high correct recognition ratios, the Levenberg-Marquardt method [18] is applied in the training process, and the value of weight coefficients in each layer can be adjusted according to the value of $E(w)$. Every adjusting process can be divided into two stages: a forth propagation and a back propagation stage. The process does not stop until the value of the evaluation function is smaller than a specific value, and the rest weight values between different layers is the neural network which is what we need. In this study, MATLAB neural network toolbox is used as the research platform [19].

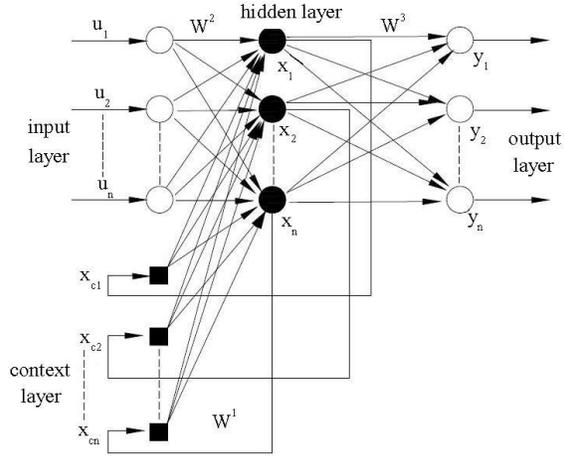

Fig. 1. The structure of the ENN.

## 3. Experiment

### 3.1 Experiment setup

The experimental setup is similar to that which has been presented in detail in Yu et al. [13]. A quartz glass vessel with dimensions of 30 cm in diameter and 40 cm in length is filled with EJ-335 LS loaded with 0.5 percent gadolinium (Gd) by weight [20]. Two 20 cm Hamamatsu R5912-02 photomultiplier tubes (PMT) [21] are mounted on each side of the cylindrical vessel. In order to enhance the light collection efficiency, Plexiglas light guides are set to fit the surface shapes of the cylindrical vessel and PMTs, and the interfaces of these parts are coupled by silicone grease. In the experiment, random triggers are applied to measure the dead time of the detector. The signals from PMT1, PMT2 and random signals generated by a TEKTRONIX AFG3000C function signal generator are all separately sent to a CAEN N625 Fan-in fan-out (FIFO) module which divides each signal into two circuits. One part of these three signals are sent to a CAEN V1721 fast analog to digital conversion (FADC) module with 8 bit resolution working at 500 M sample/s, and the other part of these three signals are sent to a CAEN N842 discriminator which can discriminate the signals according to the maximum amplitude. The output signals of the N842 are transferred into a CAEN N405 logical module to generate a level signal. The level signal and the output signal of the random trigger after discrimination are input into logic OR to get another level signal to trigger the V1721 FADC. The data of the V1721 are collected by a computer via the CAEN A2818 photoelectric conversion module. The detector is placed in a lead chamber with 5 cm thickness to shield the environment γ background. The experimental setup is shown in Fig. 2.

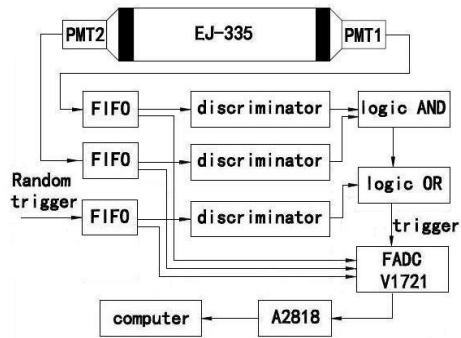

Fig. 2.  Electronics and data acquisition system diagram of the LS fast neutron detector.

The entire set-up is exposed to an $^{241}$Am–$^9$Be neutron source. The continuous neutron spectrum of a conventional $^{241}$Am–$^9$Be source, giving rise to an average energy of 4.5 MeV with a range of 0-10 MeV, is employed.  The source is set in the middle of one side and 130 cm to the center of the detector.

Incident neutrons will be slowed down by colliding with nuclei in the LS, and transfer their energies to these recoiled nuclei. Due to the energy loss, the primary neutrons will be turned into thermal neutrons, while the recoiled nuclei excite or ionize atoms in the LS, and generate light pulses. The remaining thermal neutrons will be captured by the Gd nucleus in their trajectory terminus and generate excited nuclei, which eventually emit several γ rays on the time-scale of nanoseconds with a total energy of approximately 8 MeV. These γ rays could induce a light pulse via ionization of electrons generated by the photoelectric effect, Compton scatting, and pair production process. The overall process is schematically shown in Fig. 3

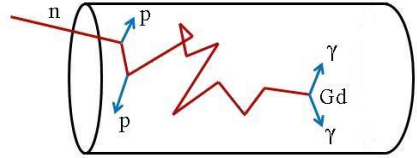

Fig. 3. (color online) Neutron slowing and capture process in LS.

### 3.2 Data Acquisition

In this study, identified neutron and γ ray signals were chosen from the scatter plot by CCM to constitute a reliable training and test data set. Two important parameters, Qpart and Qtotal, were used to describe a signal in CCM. Qtotal is the integral of the pulses spanning from 20 bins before the signal peak to 80 bins after the peak.  Because the difference of the pulse shape caused by neutrons and γs is mainly concentrated in the falling edge, the falling edge of the signal was chosen to calculate Qpart, and its value was obtained through integrating the falling edge from 15 bins to 80 bins after the peak. The discrimination factor is defined as Dis = Qpart/Qtotal. Here, the PMT1 signals were selected to perform n/γ discrimination. According to the scatter plot of CCM, we can learn that neutron and γ events with energy greater than 0.8 MeV can be separated with high confidence.  But due to the noise and fluctuations in the pulse generation, it is difficult to distinguish neutron and γ ray events around the vicinity of the gap between the two bands in which the events with energy exceed 0.8 MeV. In order to get a reliable training and test data set, we only select those of the neutron and γevents with energy exceeding 0.8 MeV from the bands marked with red and blue respectively in Fig. 4.

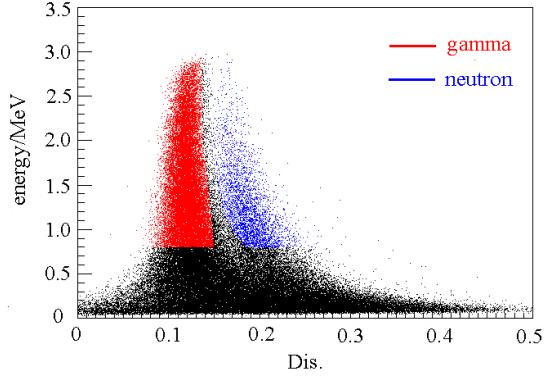

Fig. 4.  (color online) The scatter plot of the Charge Comparison Method (CCM). The x-axis is the discrimination factor Dis, and the y-axis is the energy.

These selected signals were normalized to eliminate the impacts of the difference of the pulse amplitude. The normalization process is shown in Eq. (5):

$$x' = \frac{X - X_{\min}}{X_{\max} - X_{\min}}, \qquad (5)$$

where $x$ is the pulse amplitude of each bin of a signal, $x_{max}$ is the maximum pulse amplitude, $x_{min}$ is the minimum pulse amplitude, $x'$ is the normalized pulse amplitude. Fig. 5 shows the average signals of neutrons and γ rays generated in LS. Since the falling edges of neutron and γ pulse are significantly different, we chose the 80 bins after the peak which can sufficiently describe the falling edge of pulses.

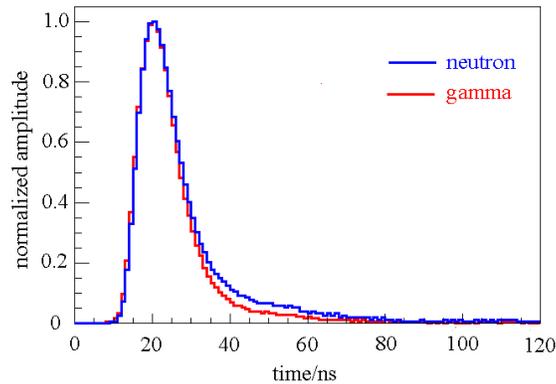

Fig. 5. (color online) The pulse time spectrum of average neutron and γ signals in EJ-335. The x-axis is time in ns, the y-axis is the amplitude of the signal.

Then the 80 normalized falling pulse time spectra were fitted with a double exponential function expressed with Eq. (6) [22] by CERN ROOT software [23]

$$f(x) = exp(p0 + p1t) + exp(p2 + p3t), \quad (6)$$

where $p0, p1, p2, p3$ are four free parameters of the double exponential function and $t$ is the time scale of a pulse. The average value of the four parameters were obtained from the selected identified 200 neutrons and 200 γ events, respectively. Then the neutron and γ training data we selected were fitted with neutron and γ fitted functions respectively. We use chi squared ($\chi^2$) as defined in Eq. (7) to evaluate the difference between the measured pulse and the fitted neutron or γ function, and select relatively smooth neutron and γ pulses as training data.

$$\chi^2 = \sum_i \frac{(F_i - f_i)^2}{f_i}, \quad (7)$$

where $F_i$ is the i-th bin value of the neutron or γ pulse shape from the detector, and $f_i$ is the i-th value of neutron or γ fitted function. The chi squares value range is 0-0.82 for γ events and 0-1.25 for neutron events. In this study, we selected the training pulses with the $\chi^2$ value less than 0.12, since most of the neutron and γ pulses are relatively smooth in this region.

## 4. Results and Discussion

506 neutron events and 648 γ ray events, after being fitted with Eq. (6), were selected as training samples for the neural network. The training data set is shown in Fig. 6, and the test data is from the red and blue band shown in Fig. 4. We chose the 80 bins after the peak of pulses to construct an 1154×80 matrix as the neural network training data set. After some tests, the hidden layer unit was set to 20. The node number of the output layer was set to 1. The ENN is thus an $80 \times 20 \times 20 \times 1$ multi-layer network. Following the forth propagation and the back propagation, the weights of ENN are adjusted in the training process. If the output value of $E(w)$ is less than 0.01 the training will be ended. Here we set the output of the ENN to be 0 or 1 according to the input event of a neutron event or a γ-ray event, respectively. After the training process is finished, the test pulse data are sent to the ENN to classify into neutron or γ events. If the output of the ENN is greater than 0.5, the event is a γ-ray event, otherwise, it is a neutron event.

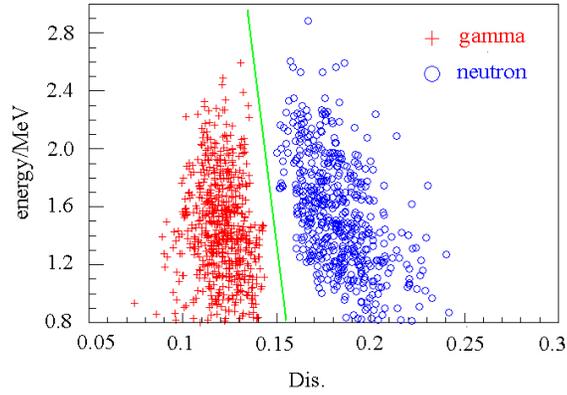

Fig. 6.  (color online) The neutron and γ training data after being fitted by double exponential function.

To evaluate the discrimination ability of the ENN, the discrimination error ratio (DER) is defined as the ratio of the number of events incorrectly discriminated by the ENN to the total number of events of the test data set. The DER of neutrons and γ rays is calculated by the following equations:

$$DER_\gamma = \frac{N_\gamma - N_{\gamma\_ENN}}{N_\gamma} \times 100\%, \qquad (8)$$

$$DER_n = \frac{N_n - N_{n\_ENN}}{N_n} \times 100\%, \qquad (9)$$

where $N_\gamma$ and $N_n$ are the test numbers of γ rays and neutrons, respectively, and $N_{\gamma\_ENN}$ and $N_{n\_ENN}$ mean the numbers of γ ray and neutron events which have been correctly classified.

Table 1.  Discrimination results of ENN

| Particle | γ | neutron |
|---|---|---|
| Train number | 648 | 506 |
| Verification number | 647 | 505 |
| Test number | 3440 | 1165 |
| Test result | 3435 | 1154 |
| DER_test(%) | 0.15 | 0.94 |

In Table 1, since the values of the DER test are really small, it is reasonable to believe that the ENN is excellent at discriminating n/γ events.

Figure 7 shows the ENN n/γ discrimination result mapped into a scatter plot, in which neutron and γ events with energy above 0.8 MeV can be separated exactly. Figure 8 is the corresponding distribution histogram of Fig. 7.

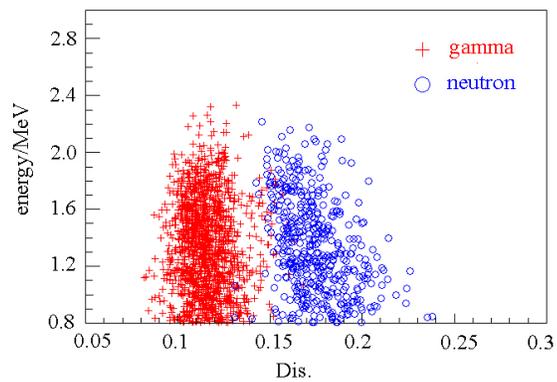

Fig. 7.  (color online) The discrimination scatter plot of ENN.

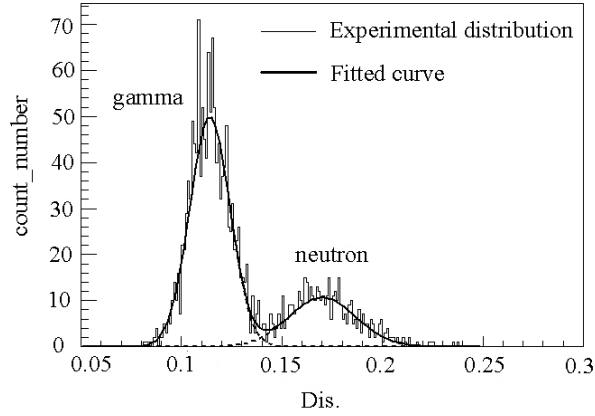

Fig. 8.  Distribution histogram corresponding to Fig. 7.

Another way to quantitatively estimate the n/γ discrimination performance of the above two methods is the Figure of Merit (FOM) [24]. A larger value of FOM indicates better performance of the n/γ discrimination. The FOM is defined as Eq. 10:

$$FOM = \frac{S}{FWHM_n + FWHM_\gamma}, \quad (10)$$

where $S$ is the interval of the peaks of the neutron and γ-ray events, $FWHM_\gamma$ is the full-width-half-maximum (FWHM) of the distribution of the events classified as γ-rays and $FWHM_n$ is the $FWHM$ of the distribution of the neutron events. If the spectra are consistent with Gaussian distributions, the FOM can be written as Eq. 11 [25]:

$$FOM = \frac{|\mu_n - \mu_\gamma|}{2.35 \cdot (\delta_n + \delta_\gamma)}, \quad (11)$$

where $\mu_n$ and $\mu_\gamma$ are the mean values of the neutron and γ distributions, respectively, and $\delta_n$ and $\delta_\gamma$ are the corresponding standard deviations of the distributions.

The fitting method is also used to choose neutron and γ pulse data in the study of FOM. A total of 1838 neutron and γ events with energy above 0.8 MeV were selected to evaluate the FOM value of ENN and BPNN, with $\chi^2$ values of neutron and γ fitted functions all less than 0.135. The comparison between FOMs of the ENN and BPNN is reported in Table 2. The results indicate that the ENN has better n/γ discrimination ability.

Table 2. Comparison of the FOM Value of ENN and BPNN

| Methods | FOM |
|---------|-----|
| ENN | 0.953±0.037 |
| BPNN | 0.907±0.034 |

## 5. Conclusion

A new n/γ digital discrimination method based on an ENN was presented. To satisfy the requirements of machine learning for the method, the CCM and fitting method were utilized to acquire the training data set and the test data set from signals above 0.8 MeV of an EJ-335 LS detector. The discrimination performance of the ENN was evaluated through comparison with BPNN method. Both networks were properly trained and tested with the above training and test data set. The results indicate that the ENN method performed better than the BPNN method. Compared to the BPNN, which had a FOM of 0.907 ± 0.034, the FOM of the ENN was 0.953 ± 0.037, which indicates the ENN has the capability of effective n/γ discrimination. Since it stores the internal status and mapping dynamic characteristics in a recognition model, the classification ability of the ENN is more stable and robust. In conclusion, the ENN discrimination method provides a new choice of n/γ pulse shape discrimination which uses pattern recognition.

**Reference:**

1  V. Chazal, R. Brissot, J. F. Cavaignac et al, Astropart. Phys., **9**: 163 (1998)

2  H. Y. Xing, L. Wang, J. J. Zhu et al, Chin. Phys. C, **37**: 026003 (2013)

3  H. Wulandari, J. Jochum, W. Rau et al, Astropart. Phys., **22**: 313 (2004)

4  D. Jordan., J. L. Tain., A. Algora. et al, Astropart. Phys., **42**: 1 (2013)

5  D. Z. Ding, C. Y. Ye, Z. X. Zhao et al, *Neutron Physics* (Beijing, Atomic Energy Press, 2001), p. 132 (in Chinese)

6  F. D. Brooks, Nucl. Instrum. Methods, **4**: 151 (1959).

7  J. M. Adams, G.White, Nucl. Instrum. Methods, **156**: 459 (1978)

8  T. K. Alexander, F. S. Goulding, Nucl. Instrum. Methods, **13**: 244 (1961)

9  M. L. Roush, M. A. Wilson, W. F. Hornyak, Nucl. Instrum. Methods, **31**: 112 (1964)

10  D. Savran, B. Loher, M. Miklacec et al, Nucl. Instrum. Methods. A, **624**: 675 (2010)

11  S. Yousefi, L. Lucchese, M. D. Aspinall, Nucl. Instrum. Methods. A, **598**: 551 (2009)

12  X. L. Luo, Y. K. Wang, J. Yang et al, Nucl. Instrum. Methods. A, **717**: 44 (2013)

13  X. Z. Yu, J. J. Zhu, S. T. Lin et al, Nucl. Instrum. Methods. A, **777**: 80 (2015)

14  C. Zhong, L. F. Miller, M. Buckner, Nucl. Instrum. Methods. A, **416**: 438 (1998)

15  B. Esposito, L. Fortuna, A. Rizzo, Neural neutron/gamma discrimination in organic scintillators for fusion application, *In Proceeding of the 2004 IEEE International joint Conference on Neural networks*, edited by IEEE (USA: IEEE, 2004), p. **4**: 2931

16  G. Liu, M. D. Aspinall, X. Ma et al, Nucl. Instrum. Methods. A, **607**: 620 (2009)

17  J. L. Elman, Cogn. Sci., **14**: 179 (1990)

18  B. G. Kermani, S. S. Schiffman, H. T. Nagle et al. Sens. Actuators, B, **110**: 13 (2005)

19  H. Demuth, M. Beale, M. Hagan, http://kashanu.ac.ir/Files/Content/neural_network_toolbox_6.pdf, retrieved 1998

20  EJ-331 and EJ-335 Gadolinium Loaded Liquid Scintillator, http://www.ggg-tech.co.jp/maker/eljen/ej-331.html

21  HAMAMATSU, http://www.hamamatsu.com/resources/pdf/etd/LARGE_AREA_PMT_TPMH1286E05.pdf

22  S. Marrone, D. Cano-Ott, N. Colonna et al, Nucl. Instrum. Methods. A, **490**: 299 (2002)

23  I. Antcheva, M. Ballintijn, B. Bellenot et al, Comput. Phys. Commun, **182**: 1384 (2011)

24  R. A. Winyard, J. E. Lukin, G. W. Mcbeth, Nucl. Instrum. Methods, **95**: 141 (1971)

25  G. Liu, M. J. Joyce, X. Ma et al, IEEE Trans. Nucl. Sci., **57**: 1682 (2010)